# Grey Level Co-occurrence Matrix (GLCM) Based Second Order Statistics for Image Texture Analysis

Abdul Rasak Zubair[1], Oluwaseun Adewunmi Alo[2]
[1,2]Department of Electrical and Electronic Engineering, University of Ibadan, Ibadan, Nigeria
([1]ar.zubair@ui.edu.ng, [2]oluwaseunalo@gmail.com)

*Abstract*- Grey Level Co-occurrence Matrix (GLCM) and Grey Level Difference Vector (GLDV) are described and computed for twenty four 128 x 128 x 3 test images along horizontal, vertical and diagonal directions. Second order image statistics such as Contrast, Dissimilarity, Homogeneity (Inverse Difference Moment), Angular Second Moment (ASM), Energy, Maximum Probability, Entropy, Mean (μ), Standard Deviation (σ) and Correlation are computed and studied. GLDV gives the Probabilities of Occurrence of Difference of 0, 1, 2, 3, … , 254, 255. Group GLDV gives the thirteen (13) Probabilities of Occurrence of Difference of 0-19, 20-39, 40-59, … , 220-239, 240-255 which can be displayed with bar charts. The results show that smooth images have lower Contrast values and higher Probability of Occurrence of Difference of 0 – 19 while rough images have higher Contrast values and lower Probability of Occurrence of Difference of 0 – 19. The degree of smoothness or roughness of an image may not be exactly the same along horizontal, vertical and diagonal directions. There are significant correlation between Dissimilarity & Contrast, Homogeneity & Contrast, Entropy & Contrast, Energy & Contrast, Standard Deviation (σ) & Contrast, Correlation & Contrast, and Probability of Occurrence of Difference of 0 – 19 & Contrast with correlation coefficients of +0.9322, -0.5011, 0.6681, -0.4255, -0.4914, -0.5428, and -0.8346 respectively.

## I. INTRODUCTION

Common texture terms are rough, smooth or silky and bumpy. These refer to touch. Texture is connected with changes in elevation between the high and the low points on a topographical surface. Rough means large difference between the high and low points. Silky or smooth means little difference between the high and low points. Image texture refers to changes in brightness values (Grey levels) and not changes in elevation. Image texture analysis are necessary in medical diagnosis, image processing and segmentation, remote sensing, biological and chemical sciences [1, 2, 3, 4, 5, 6, 7, 8].

Grey Level Co-occurrence Matrix (GLCM) is formed from an Image. The descriptive statistics derived from GLCM are important image texture measures. These descriptive statistics are known as second order statistics [9, 10, 11, 12, 13]. First order statistics are those derived from the image itself [14, 15].

Given the sample Test Image of Fig. 1, the numbers of co-occurrences of pairs of grey values are recorded in the GLCM. A pair is made of reference pixel i and neighbor pixel j. There are eight possible directions. For example, the $3^{rd}$ row / $3^{rd}$ Column pixel with grey level 4 as a reference pixel has eight possible neighbors. The co-occurrence grey levels are (4,6), (4,0), (4,1), (4,0), (4,1), (4,0), (4,3), and (4,2) along $0^o$ (East), $45^o$ (North East), $90^o$ (North), $135^o$ (North West), $180^o$ (West), $225^o$ (South West), $270^o$ (South), and $315^o$ (South East) respectively. Therefore, there are eight possible GLCM that can be formed for the Test Image.

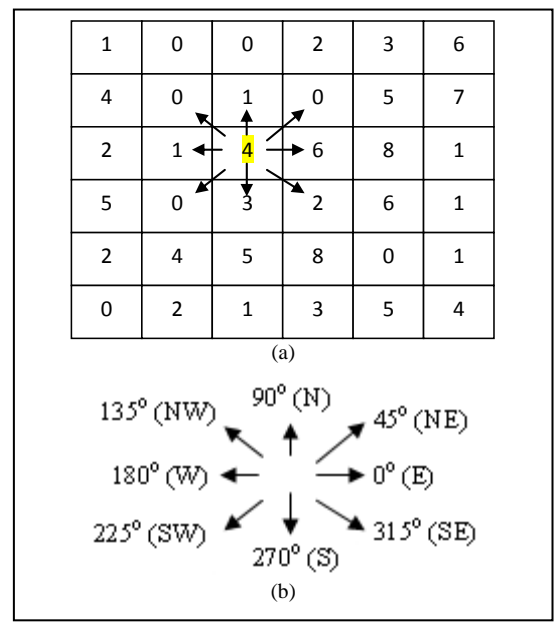

Figure 1. a) Test Image Pixels, b) 8 Directions

These are reduced to three possible symmetric GLCM; namely Horizontal GLCM [addition of $0^o$ (East) GLCM and $180^o$ (West) GLCM], Vertical GLCM [addition of $90^o$ (North) GLCM and $270^o$ (South) GLCM], and Diagonal GLCM [addition of $45^o$ (North East) GLCM and $225^o$ (South West) GLCM] or [addition of $135^o$ (North West) GLCM and $315^o$ (South East) GLCM]. Both diagonals give the same Diagonal GLCM.

## II. GREY LEVEL CO-OCCURRENCE MATRIX (GLCM)

### A. Horizontal Grey Level Co-occurrence Matrix (GLCM)

The ($0^o$) East Grey Level Co-occurrence Matrix (GLCM) is formed from Fig. 1 as shown in Fig. 2. Pixels in the last column of Fig. 1 cannot serve as reference pixels as they do not have Eastern neighbor. Each of other pixels in Fig. 1 serves as reference pixel value i and its corresponding Eastern neighbor pixel value j are noted as pair (i,j). For example, (1,0), (0,0), (0,2), (2,3), and (3,6) are five (5) pairs noticed on the first row of Fig. 1. There are thirty (30) pairs for the Test Image of Fig. 1. If the test image is an rgb colour image, the image would have three 2D matrices and there would be 90 pairs. The numbers of co-occurrences of pairs of pixel values are recorded in the shaded GLCM of Fig. 2. Row 1 (i=0) / column 1 (j=0) box contain the number of times (0,0) occurs. Row 7 (i=6) / column 9 (j=0) box contains the number of times (6,8) occurs. (4,0) occurs twice, (5,8) occurs once, (1,4) occurs twice and (8,8) occurs 0 times just to give some examples. A computer program can handle this task. The pixel value in the sample Test Image varies from 0 to 8, hence the GLCM of Fig. 2 is a 9 by 9 matrix. In reality, pixel value varies from 0 to 255, therefore, GLCM is actually a 256 by 256 matrix.

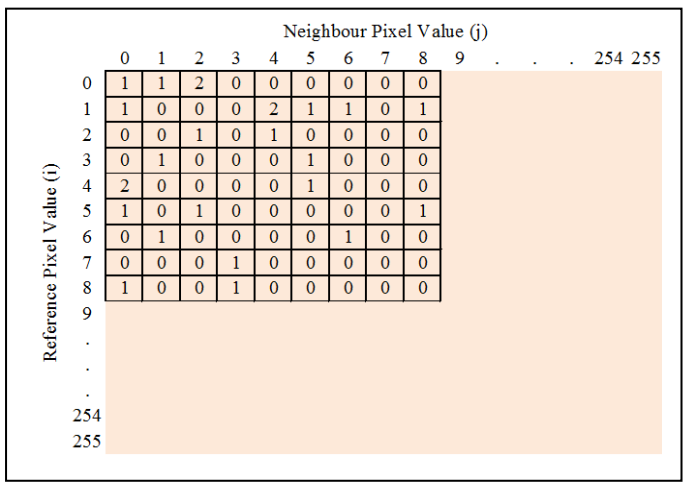

Figure 2. ($0^o$) East GLCM.

(0,1), (0,0), (2,0), (3,2), and (6,3) are five (5) pairs noticed on the first row of Fig. 1 when ($180^o$) West direction is considered. The transpose of the ($0^o$) East GLCM gives the ($180^o$) West GLCM. Addition of the ($0^o$) East GLCM and the ($180^o$) West GLCM gives the Horizontal GLCM of Fig. 3. The Horizontal GLCM is symmetrical around its diagonal.

### B. Vertical Grey Level Co-occurrence Matrix (GLCM)

The ($180^o$) South Grey Level Co-occurrence Matrix (GLCM) is formed from Fig. 1 as shown in Fig. 4. Pixels in the last row of Fig. 1 cannot serve as reference pixels as they do not have Southern neighbor. Each of the other pixels in Fig. 1 serves as reference pixel value i and its corresponding Southern neighbor pixel value j are noted as pair (i,j). For example, (1,4), (4,2), (2,5), (5,2), and (2,0) are five (5) pairs noticed on the first column of Fig. 1. Fig. 4 is the ($270^o$) South GLCM.

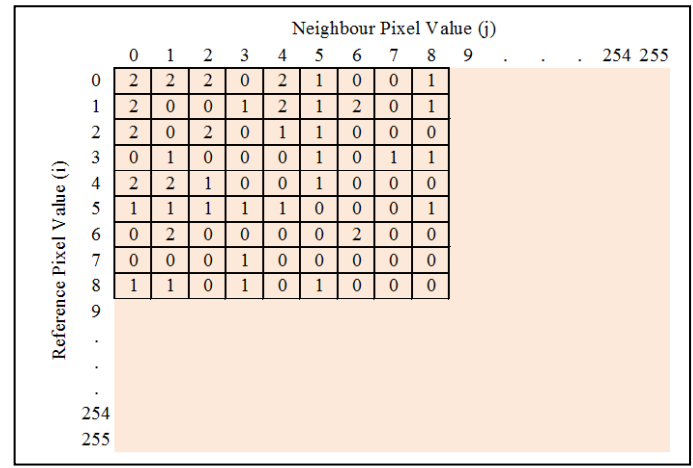

Figure 3. Horizontal GLCM.

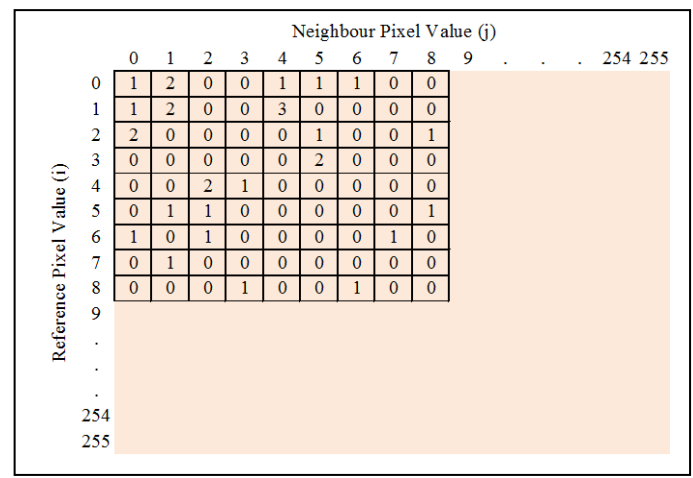

Figure 4. ($270^o$) South GLCM.

(4,1), (2,4), (5,2), (2,5), and (0,2) are five (5) pairs noticed on the first column of Fig. 1 when ($90^o$) North direction is considered. The transpose of the ($270^o$) South GLCM gives the ($90^o$) North GLCM. Addition of the ($270^o$) South GLCM and the ($90^o$) North GLCM gives the Vertical GLCM of Fig. 5. The Vertical GLCM is also symmetrical around its diagonal.

### C. Diagonal Grey Level Co-occurrence Matrix (GLCM)

The ($315^o$) South East Grey Level Co-occurrence Matrix (GLCM) is formed from Fig. 1 as shown in Fig. 6. Pixels in the last row of Fig. 1 cannot serve as reference pixels as they do not have South-Eastern neighbor. Each of the other pixels in Fig. 1 serves as reference pixel value i and its corresponding

Southern neighbor pixel value j are noted as pair (i,j). For example, (1,0), (4,1), (2,0), (5,4), and (2,2) are five (5) pairs noticed with reference pixels in the first column of Fig. 1. Fig. 6 is the (270°) South East GLCM.

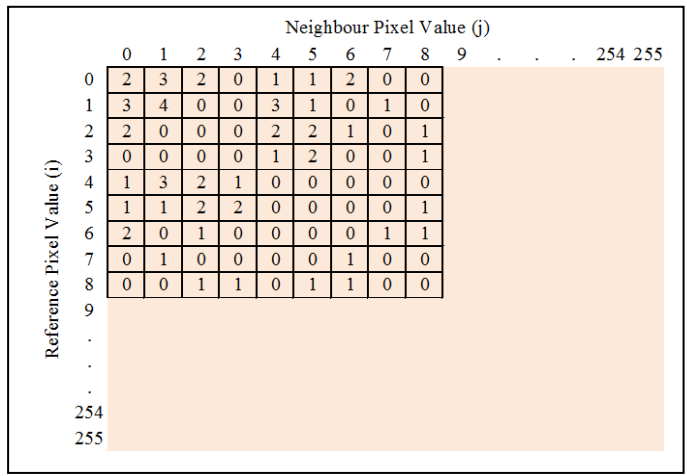

Figure 5.  Vertical GLCM.

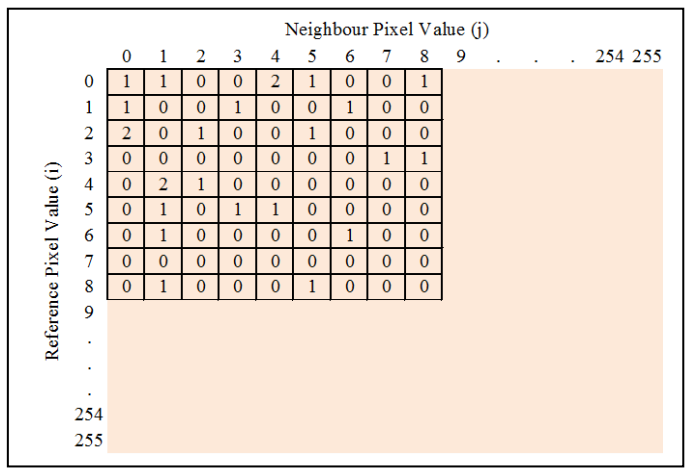

Figure 6.  (315°) South East GLCM.

(0,1), (1,4,), (0,2), (4,5), and (2,2) are five (5) pairs noticed with reference pixels in the second column of Fig. 1 when (90°) North West direction is considered. The transpose of the (315°) South East GLCM gives the (135°) North West GLCM. Addition of the (315°) South East GLCM and the (135°) North West GLCM gives the Diagonal GLCM of Fig. 7. Addition of the (45°) North East GLCM and the (225) South West GLCM gives the same Diagonal GLCM of Fig. 7. The Diagonal GLCM is also symmetrical around its diagonal.

### D. Normalized Grey Level Co-occurrence Matrix (NGLCM)

Normalized GLCM is obtained by dividing each element in GLCM with the sum of all elements in GLCM as described by Eqn. (1). This is applicable to the three distinct types of GLCM; Horizontal GLCM, Vertical GLCM and Diagonal GLCM. NGLCM (i,j) is the probability P(i,j) of Co-occurrence of the pair (i,j). Fig, 8 shows Diagonal NGLCM.

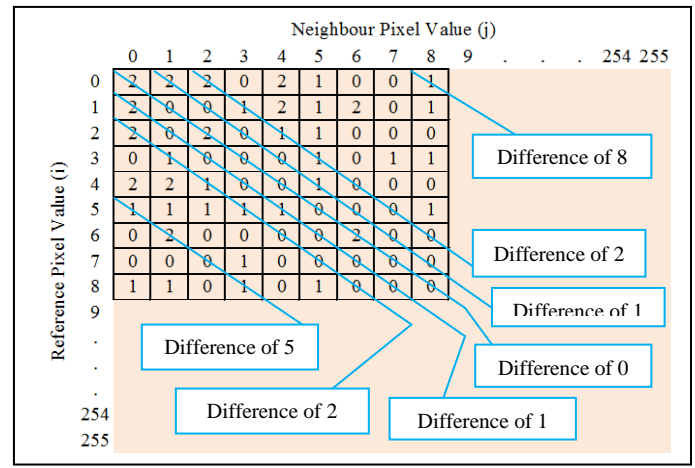

Figure 7.  Diagonal GLCM.

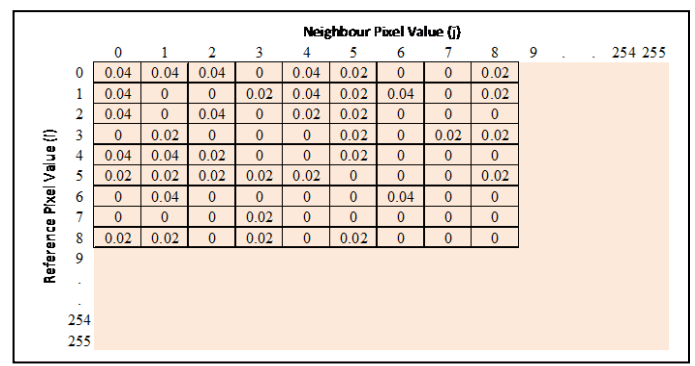

Figure 8.  Diagonal NGLCM (Probability of Co-occurrence).

$$P(i,j) = NGLCM(i,j) = \frac{GLCM(i,j)}{\sum_{i=0}^{255} \sum_{j=0}^{255} GLCM(i,j)} \quad (1)$$

### E. Grey Level Difference Vector (GLDV)

A pair (i,j) has a Difference of $|i - j|$ between the Reference pixel value and the Neighbor pixel value. The sum of the elements along the diagonal of GLCM gives the number of occurrence of Difference of 0 between the Reference and Neighbor pixels' values as illustrated in Fig. 7. Sum of the elements along the diagonal of NGLCM gives the Probability of Occurrence of Difference of 0. Two lines parallel to the diagonal on the GLCM and NGLCM represent each of the Difference of 1, 2, 3, …, 255 as illustrated in Fig. 7. Fig. 9 shows the Diagonal Grey Level Difference Vector (GLDV) formed from the Diagonal GLCM of Fig. 7. Horizontal GLDV and Vertical GLDV can also be obtained from Horizontal GLCM and Vertical GLCM respectively. For the purpose of Texture analysis, GLDV which is a 256 by 3 matrix can be reduced to Group GLDV which is a 13 by 3 matrix by breaking down 0,1,2,…,254,255 into groups 0-19, 20-39, 40-59, …, 220-239, 240-255 as illustrated in Fig. 10. The number of Occurrences of Difference of 0,1,2,…, 18 and 19 in GLDV are added to give the number of Occurrences of Difference of 0-19 in Group GLDV. Bar chart can display the Group GLDV and show the nature of the Image under study.

| Difference of | Number of Occurrence | Probability of Occurrence |
|---|---|---|
| 0 | 6 | 0.1000 |
| 1 | 10 | 0.1667 |
| 2 | 14 | 0.2333 |
| 3 | 12 | 0.2000 |
| 4 | 6 | 0.1000 |
| 5 | 4 | 0.0667 |
| 6 | 8 | 0.1333 |
| 7 | 0 | 0.0000 |
| 8 | 0 | 0.0000 |
| 9 | 0 | 0.0000 |
| . | | |
| . | | |
| . | | |
| 254 | | |
| 255 | | |

Figure 9. Diagonal GLDV.

| Difference of | Number of Occurrence | Probabilty of Occurrence |
|---|---|---|
| 0 - 19 | 60 | 1.0000 |
| 20 -39 | 0 | 0.0000 |
| 40 - 59 | 0 | 0.0000 |
| 60 - 79 | 0 | 0.0000 |
| 80 -99 | 0 | 0.0000 |
| 100 - 119 | 0 | 0.0000 |
| 120 -139 | 0 | 0.0000 |
| 140 - 159 | 0 | 0.0000 |
| 160 - 179 | 0 | 0.0000 |
| 180 -199 | 0 | 0.0000 |
| 200 - 219 | 0 | 0.0000 |
| 220 -239 | 0 | 0.0000 |
| 240 - 255 | 0 | 0.0000 |

Figure 10. Diagonal Group GLDV.

## III. IMAGE TEXTURE MEASURES: SECOND ORDER STATISTICS

Second order statistics of an image are single statistical values used to summarize normalized symmetrical GLCM. They are also called image texture measures. They are different from first order statistics like Brightness, Contrast, histogram and Frequency Estimate which are derived directly from the pixel values [14, 15]. The Second order image statistics are Contrast, Dissimilarity, Homogeneity (Inverse Difference Moment), Angular Second Moment (ASM), Energy, Maximum Probability, Entropy, Mean (μ), Standard Deviation (σ) and Correlation which are given by Eqns. (2), (3), (4), (5), (6), (7), (8), (9), (10), and (11) respectively.

$$Contrast = \sum_{i=0}^{255}\sum_{j=0}^{255}(i-j)^2 P(i,j) \quad (2)$$

$$Dissimilarity = \sum_{i=0}^{255}\sum_{j=0}^{255}|i-j|P(i,j) \quad (3)$$

$$Homogeneity = \sum_{i=0}^{255}\sum_{j=0}^{255}\frac{P(i,j)}{1+(i-j)^2} \quad (4)$$

$$ASM = \sum_{i=0}^{255}\sum_{j=0}^{255}[P(i,j)]^2 \quad (5)$$

$$Energy = \sqrt{ASM} \quad (6)$$

Maximum Probability = Maximum element in NGLCM  (7)

$$Entropy = \sum_{i=0}^{255}\sum_{j=0}^{255} -P(i,j)Log_e[P(i,j)] \quad (8)$$

$$\mu = \sum_{i=0}^{255}\sum_{j=0}^{255} iP(i,j) = \sum_{i=0}^{255}\sum_{j=0}^{255} jP(i,j) \quad (9)$$

$$\sigma = \sqrt{\sum_{i=0}^{255}\sum_{j=0}^{255}(i-\mu)^2 P(i,j)} = \sqrt{\sum_{i=0}^{255}\sum_{j=0}^{255}(j-\mu)^2 P(i,j)} \quad (10)$$

$$Correlation = \sum_{i=0}^{255}\sum_{j=0}^{255} P(i,j)\frac{(i-\mu)(j-\mu)}{\sigma^2} \quad (11)$$

In this work, computer programs are developed to compute the three types of GLCM, compute their corresponding ten statistics, and plot their corresponding GLDV for image texture analysis. Twenty four 128 x 128 x 3 test images with are analyzed.

## IV. RESULTS AND DISCUSSIONS

The Contrast values and the Probability of Occurrence of Difference of 0 - 19 (the element in the first row and third column of Group GLDV) obtained for the twenty four test images based on Horizontal NGLCM, the Vertical NGLCM and the Diagonal NGLCM are presented in Table I. For each test image, the Average Contrast is computed by dividing the sum of the Contrast values obtained from the Horizontal, Vertical and Diagonal GLCM by three (3). Table II shows the twenty four test images and their average Contrast values. The smooth images tend to have lower Average Contrast values while rough images tend to have higher Average Contrast values.

TABLE I. CONTRAST AND PROBABILITY OF OCCURRENCE OF DIFFERENCE OF 0 – 19 FOR TWENTY FOUR TEST IMAGES

| Based on Horizontal NGLCM | | | Based on Vertical NGLCM | | | Based on Diagonal NGLCM | | |
|---|---|---|---|---|---|---|---|---|
| Image | Contrast | Probability of Occurrence of Difference of 0 - 19 | Image | Contrast | Probability of Occurrence of Difference of 0 - 19 | Image | Contrast | Probability of Occurrence of Difference of 0 - 19 |
| Test Image 1 | 1.11 | 1.0000 | Test Image 1 | 1.00 | 1.0000 | Test Image 1 | 1.74 | 1.0000 |
| Test Image 2 | 17.23 | 0.9934 | Test Image 2 | 21.17 | 0.9898 | Test Image 2 | 37.56 | 0.9780 |
| Test Image 3 | 25.77 | 0.9967 | Test Image 3 | 23.54 | 0.9984 | Test Image 3 | 39.36 | 0.9921 |
| Test Image 5 | 54.23 | 0.9739 | Test Image 4 | 43.74 | 0.9840 | Test Image 4 | 106.54 | 0.9343 |
| Test Image 6 | 72.11 | 0.9555 | Test Image 7 | 54.43 | 0.9836 | Test Image 5 | 108.74 | 0.9241 |
| Test Image 4 | 85.04 | 0.9534 | Test Image 6 | 68.37 | 0.9589 | Test Image 6 | 133.30 | 0.9149 |
| Test Image 7 | 136.90 | 0.9062 | Test Image 5 | 84.88 | 0.9442 | Test Image 7 | 162.23 | 0.8793 |
| Test Image 8 | 196.38 | 0.8938 | Test Image 9 | 113.32 | 0.9306 | Test Image 8 | 247.41 | 0.9316 |
| Test Image 17 | 208.29 | 0.9134 | Test Image 11 | 199.77 | 0.8465 | Test Image 10 | 384.33 | 0.7646 |
| Test Image 10 | 214.75 | 0.8610 | Test Image 8 | 201.32 | 0.8898 | Test Image 9 | 396.51 | 0.7556 |
| Test Image 14 | 234.99 | 0.9032 | Test Image 10 | 276.86 | 0.8214 | Test Image 11 | 458.82 | 0.6802 |
| Test Image 9 | 363.51 | 0.7714 | Test Image 15 | 415.17 | 0.7499 | Test Image 12 | 538.48 | 0.6273 |
| Test Image 11 | 364.36 | 0.7369 | Test Image 13 | 419.78 | 0.7704 | Test Image 13 | 734.85 | 0.6038 |
| Test Image 13 | 419.69 | 0.7703 | Test Image 12 | 483.04 | 0.6570 | Test Image 14 | 778.82 | 0.7951 |
| Test Image 12 | 506.23 | 0.6454 | Test Image 16 | 568.30 | 0.5828 | Test Image 15 | 902.35 | 0.6218 |
| Test Image 15 | 529.70 | 0.7135 | Test Image 14 | 625.61 | 0.8415 | Test Image 16 | 938.40 | 0.4096 |
| Test Image 16 | 598.39 | 0.5694 | Test Image 19 | 798.37 | 0.7942 | Test Image 22 | 1184.56 | 0.4313 |
| Test Image 20 | 686.19 | 0.6501 | Test Image 18 | 900.42 | 0.5865 | Test Image 18 | 1324.11 | 0.4642 |
| Test Image 18 | 803.89 | 0.5813 | Test Image 21 | 1125.39 | 0.4731 | Test Image 17 | 1344.91 | 0.6572 |
| Test Image 19 | 814.89 | 0.8014 | Test Image 17 | 1189.08 | 0.7036 | Test Image 20 | 1354.56 | 0.5416 |
| Test Image 23 | 1106.27 | 0.4359 | Test Image 20 | 1373.55 | 0.5321 | Test Image 19 | 1533.77 | 0.6762 |
| Test Image 21 | 1131.30 | 0.4602 | Test Image 23 | 1501.96 | 0.3676 | Test Image 21 | 1542.63 | 0.4055 |
| Test Image 22 | 1139.71 | 0.4309 | Test Image 22 | 1706.43 | 0.3525 | Test Image 23 | 1724.51 | 0.3375 |
| Test Image 24 | 3462.06 | 0.3377 | Test Image 24 | 2276.93 | 0.3984 | Test Image 24 | 3780.63 | 0.3449 |

TABLE II. THE TWENTY FOUR TEST IMAGES AND THEIR AVERAGE CONTRAST VALUES

| Image | 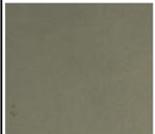 | 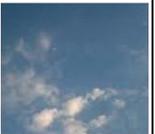 | 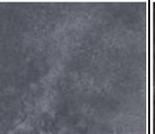 | 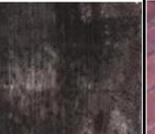 | 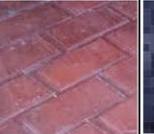 | 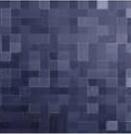 |
|---|---|---|---|---|---|---|
| Label | Test Image 1 | Test Image 2 | Test Image 3 | Test Image 4 | Test Image 5 | Test Image 6 |
| Average Contrast | 1.28 | 25.32 | 29.56 | 78.44 | 82.62 | 91.26 |
| Image | 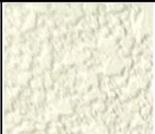 | 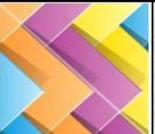 | 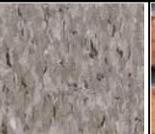 | 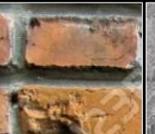 | 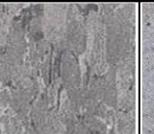 | 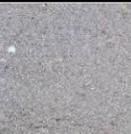 |
| Label | Test Image 7 | Test Image 8 | Test Image 9 | Test Image 10 | Test Image 11 | Test Image 12 |
| Average Contrast | 117.85 | 215.04 | 291.11 | 291.98 | 340.98 | 509.25 |
| Image | 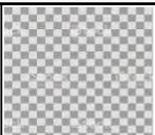 | 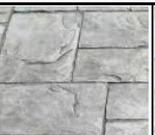 | 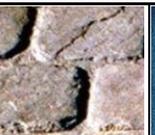 | 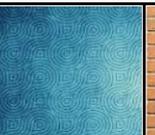 | 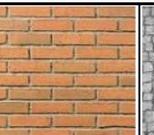 | 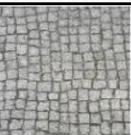 |
| Label | Test Image 13 | Test Image 14 | Test Image 15 | Test Image 16 | Test Image 17 | Test Image 18 |
| Average Contrast | 524.77 | 546.47 | 615.74 | 701.70 | 914.09 | 1009.48 |
| Image | 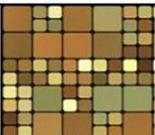 | 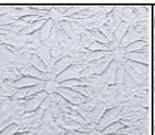 | 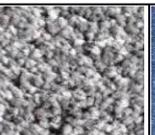 | 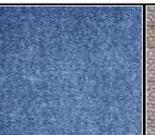 | 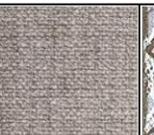 | 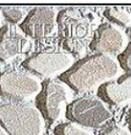 |
| Label | Test Image 19 | Test Image 20 | Test Image 21 | Test Image 22 | Test Image 23 | Test Image 24 |
| Average Contrast | 1049.01 | 1138.10 | 1266.44 | 1343.56 | 1444.25 | 3173.21 |

The actual Contrast values and the Probability of Occurrence of Difference of 0 - 19 for the twenty four test images and for the three types of GLCM are presented in graphical form in Fig. 11. Smooth images have lower Contrast values and rough images have higher Contrast values. Smooth images have higher Probability of Occurrence of Difference of 0 – 19 while rough images have lower Probability of Occurrence of Difference of 0 – 19.

Image textures as viewed along horizontal, vertical and diagonal directions are different. The degree of smoothness or roughness of an image may not be exactly the same along horizontal, vertical and diagonal directions, as observed in Table I and Fig. 11. For example, Test Image 17 has Horizontal, Vertical and Diagonal Contrast values of 208.29, 1189.08 and 1344.91 respectively as shown in Tables 1. Test Image 17 can be said to be smooth along the horizontal direction but rough along both the vertically and the diagonal directions.

The bar charts of the Horizontal Group GLDV for six test images are shown in Fig. 12. The test image for which a bar chart is plotted is shown on the bar chart. The Contrast value for the test image is also indicated. It's observed in Fig. 12 that the lower the Contrast value, the closer to 1 is the Probability of Occurrence of Difference of 0 – 19 while the closer to 0 are the Probabilities of Occurrence of Difference of 20 – 39, 40 – 59, …, and 240 – 255. As the Contrast value increases, the Probability of Occurrence of Difference of 0 – 19 reduces below 1 while the Probabilities of Occurrence of Difference of some other groups increases above 0. The number of groups with non-zero Probabilities also increases as the Contrast value increases.

The computed second order statistics and the Probability of Occurrence of Difference of 0 – 19 are presented in Tables III, IV, and V in the Appendix. The relationships between Contrast and other second order statistics are examined. There are significant correlation between Dissimilarity & Contrast, Homogeneity & Contrast , Entropy & Contrast, Energy & Contrast, Standard Deviation & Contrast, Correlation & Contrast, and Probability of Occurrence of 0 – 19 & Contrast with correlation coefficients of +0.9322, -0.5011, 0.6681,  -0.4255, -0.4914, -0.5428, and -8346 respectively.  ASM, Mean and maximum Probability are not related with Contrast. However, all the second order statistics are important for complete description of Image Texture and Image characteristics.

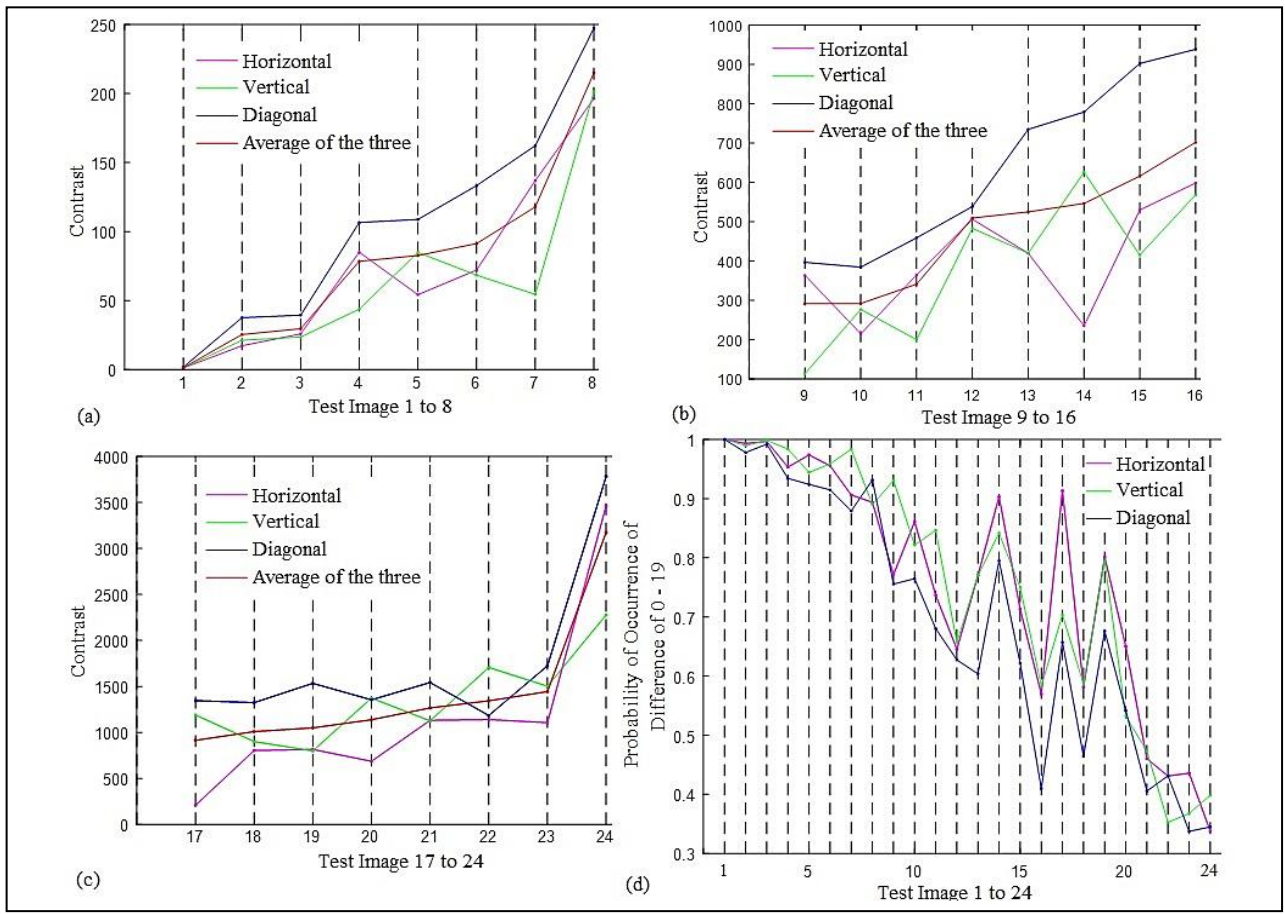

Figure 11. Graph of Contrast values and Probility of Occurrence of Difference of 0 – 19 for the three types of GLCM.

## V. CONCLUSION

The Image Texture analysis with the aid of Grey Level Co-occurrence Matrix (GLCM) has been investigated and illustrated with twenty four test images. Second order statistics and their inter-relationships have been examined. Low Contrast values and high Probability of Occurrence of Difference of 0 – 19 are associated with smooth images. High Contrast values and low Probability of Occurrence of Difference of 1 – 19 are associated with rough images. Computation of second order statistics has been made simple.

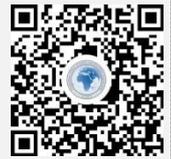

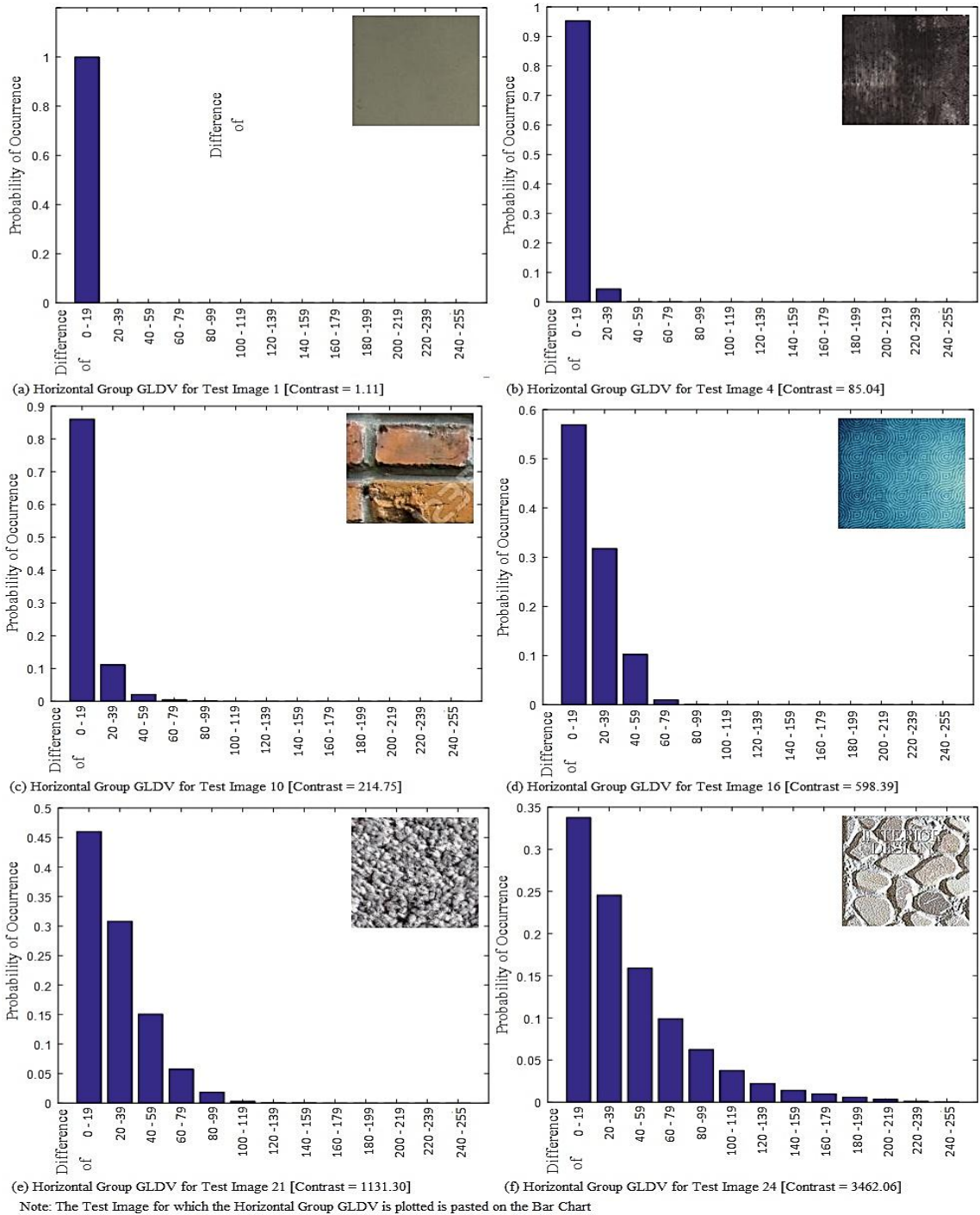

Figure 12. Bar charts of the Horizontal Group GLDV for six test images.

APPENDIX

TABLE III.    TWENTY FOUR TEST IMAGES AND THEIR SECOND STATISTICS BASED ON HORIZONTAL GLCM

| Image | Contrast | Dissimilarity | Homogeneity (Inverse Difference Moment) | Angular Second Moment (ASM) | Entropy | Mean (μ) | Energy | Standard Deviation (σ) | Correlation | Maximum Probability P(i,j) | Probability of Occurrence of Difference of 0 - 19 |
|---|---|---|---|---|---|---|---|---|---|---|---|
| Test Image 1 | 1.11 | 0.58 | 0.7531 | 0.0143 | 4.76 | 111.88 | 0.1197 | 10.47 | 0.9950 | 0.0379 | 1.0000 |
| Test Image 2 | 17.23 | 2.12 | 0.5524 | 0.0022 | 6.95 | 116.07 | 0.0465 | 30.70 | 0.9909 | 0.0080 | 0.9934 |
| Test Image 3 | 25.77 | 3.74 | 0.2528 | 0.0018 | 6.71 | 100.38 | 0.0419 | 10.71 | 0.8878 | 0.0042 | 0.9967 |
| Test Image 5 | 54.23 | 4.89 | 0.2674 | 0.0006 | 7.89 | 119.60 | 0.0251 | 28.70 | 0.9671 | 0.0029 | 0.9739 |
| Test Image 6 | 72.11 | 4.40 | 0.3882 | 0.0012 | 7.51 | 76.85 | 0.0341 | 25.56 | 0.9448 | 0.0049 | 0.9555 |
| Test Image 4 | 85.04 | 6.71 | 0.1583 | 0.0008 | 7.63 | 58.07 | 0.0276 | 18.10 | 0.8702 | 0.0019 | 0.9534 |
| Test Image 7 | 136.90 | 8.85 | 0.1188 | 0.0005 | 7.89 | 222.50 | 0.0229 | 16.81 | 0.7579 | 0.0033 | 0.9062 |
| Test Image 8 | 196.38 | 7.28 | 0.3022 | 0.0013 | 8.46 | 166.23 | 0.0367 | 60.47 | 0.9732 | 0.0306 | 0.8938 |
| Test Image 17 | 208.29 | 8.05 | 0.1847 | 0.0003 | 8.66 | 139.23 | 0.0171 | 47.91 | 0.9546 | 0.0012 | 0.9134 |
| Test Image 10 | 214.75 | 9.55 | 0.1579 | 0.0002 | 9.05 | 114.24 | 0.0134 | 49.97 | 0.9570 | 0.0017 | 0.8610 |
| Test Image 14 | 234.99 | 8.80 | 0.1617 | 0.0006 | 8.01 | 172.12 | 0.0247 | 26.46 | 0.8322 | 0.0025 | 0.9032 |
| Test Image 9 | 363.51 | 13.63 | 0.0883 | 0.0003 | 8.41 | 138.34 | 0.0185 | 20.06 | 0.5482 | 0.0010 | 0.7714 |
| Test Image 11 | 364.36 | 14.40 | 0.0738 | 0.0003 | 8.49 | 137.38 | 0.0174 | 19.49 | 0.5205 | 0.0009 | 0.7369 |
| Test Image 13 | 419.69 | 13.63 | 0.1153 | 0.0004 | 8.28 | 187.13 | 0.0211 | 29.85 | 0.7645 | 0.0021 | 0.7703 |
| Test Image 12 | 506.23 | 17.40 | 0.0596 | 0.0003 | 8.39 | 160.71 | 0.0182 | 17.33 | 0.1573 | 0.0009 | 0.6454 |
| Test Image 15 | 529.70 | 16.21 | 0.0764 | 0.0002 | 9.27 | 150.19 | 0.0127 | 52.36 | 0.9034 | 0.0040 | 0.7135 |
| Test Image 16 | 598.39 | 19.55 | 0.0527 | 0.0001 | 9.68 | 120.08 | 0.0087 | 50.92 | 0.8846 | 0.0003 | 0.5694 |
| Test Image 20 | 686.19 | 18.72 | 0.0675 | 0.0002 | 8.86 | 202.35 | 0.0155 | 27.18 | 0.5355 | 0.0010 | 0.6501 |
| Test Image 18 | 803.89 | 21.28 | 0.0526 | 0.0001 | 9.15 | 153.29 | 0.0122 | 31.06 | 0.5834 | 0.0006 | 0.5813 |
| Test Image 19 | 814.89 | 14.61 | 0.2682 | 0.0005 | 8.86 | 107.70 | 0.0229 | 55.07 | 0.8657 | 0.0094 | 0.8014 |
| Test Image 23 | 1106.27 | 26.69 | 0.0372 | 0.0001 | 9.57 | 148.41 | 0.0093 | 31.82 | 0.4536 | 0.0002 | 0.4359 |
| Test Image 21 | 1131.30 | 26.38 | 0.0393 | 0.0001 | 9.90 | 140.95 | 0.0079 | 48.22 | 0.7567 | 0.0002 | 0.4602 |
| Test Image 22 | 1139.71 | 27.05 | 0.0349 | 0.0001 | 9.71 | 117.79 | 0.0086 | 39.14 | 0.6279 | 0.0002 | 0.4309 |
| Test Image 24 | 3462.06 | 43.27 | 0.0370 | 0.0001 | 9.99 | 174.98 | 0.0121 | 57.15 | 0.4699 | 0.0081 | 0.3377 |

TABLE IV.  TWENTY FOUR TEST IMAGES AND THEIR SECOND STATISTICS BASED ON VERTICAL GLCM

| Image | Contrast | Dissimilarity | Homogeneity (Inverse Difference Moment) | Angular Second Moment (ASM) | Entropy | Mean (μ) | Energy | Standard Deviation (σ) | Correlation | Maximum Probability P(i,j) | Probability of Occurrence of Difference of 0 - 19 |
|---|---|---|---|---|---|---|---|---|---|---|---|
| Test Image 1 | 1.00 | 0.51 | 0.7853 | 0.0162 | 4.65 | 111.88 | 0.1271 | 10.47 | 0.9954 | 0.0402 | 1.0000 |
| Test Image 2 | 21.17 | 2.30 | 0.5318 | 0.0020 | 7.03 | 116.05 | 0.0444 | 30.60 | 0.9887 | 0.0081 | 0.9898 |
| Test Image 3 | 23.54 | 3.59 | 0.2634 | 0.0018 | 6.68 | 100.39 | 0.0427 | 10.73 | 0.8977 | 0.0046 | 0.9984 |
| Test Image 4 | 43.74 | 4.59 | 0.2413 | 0.0011 | 7.30 | 58.14 | 0.0337 | 18.10 | 0.9332 | 0.0032 | 0.9840 |
| Test Image 7 | 54.43 | 5.60 | 0.1804 | 0.0008 | 7.49 | 222.52 | 0.0284 | 16.82 | 0.9038 | 0.0071 | 0.9836 |
| Test Image 6 | 68.37 | 4.32 | 0.3868 | 0.0011 | 7.51 | 76.96 | 0.0339 | 25.56 | 0.9477 | 0.0049 | 0.9589 |
| Test Image 5 | 84.88 | 6.10 | 0.2387 | 0.0005 | 8.08 | 119.62 | 0.0232 | 28.70 | 0.9485 | 0.0026 | 0.9442 |
| Test Image 9 | 113.32 | 7.46 | 0.1533 | 0.0006 | 7.93 | 138.30 | 0.0240 | 20.06 | 0.8592 | 0.0016 | 0.9306 |
| Test Image 11 | 199.77 | 10.77 | 0.0950 | 0.0004 | 8.28 | 137.41 | 0.0193 | 19.49 | 0.7369 | 0.0011 | 0.8465 |
| Test Image 8 | 201.32 | 7.22 | 0.3705 | 0.0016 | 8.36 | 166.29 | 0.0395 | 60.51 | 0.9725 | 0.0312 | 0.8898 |
| Test Image 10 | 276.86 | 11.44 | 0.1158 | 0.0001 | 9.23 | 114.77 | 0.0118 | 49.85 | 0.9443 | 0.0013 | 0.8214 |
| Test Image 15 | 415.17 | 14.45 | 0.0885 | 0.0002 | 9.19 | 150.23 | 0.0137 | 52.39 | 0.9244 | 0.0058 | 0.7499 |
| Test Image 13 | 419.78 | 13.80 | 0.1041 | 0.0004 | 8.29 | 187.14 | 0.0209 | 29.86 | 0.7646 | 0.0017 | 0.7704 |
| Test Image 12 | 483.04 | 16.96 | 0.0621 | 0.0003 | 8.38 | 160.77 | 0.0183 | 17.30 | 0.1926 | 0.0009 | 0.6570 |
| Test Image 16 | 568.30 | 18.96 | 0.0559 | 0.0001 | 9.66 | 120.15 | 0.0088 | 50.98 | 0.8907 | 0.0002 | 0.5828 |
| Test Image 14 | 625.61 | 13.15 | 0.1339 | 0.0005 | 8.20 | 172.11 | 0.0227 | 26.50 | 0.5546 | 0.0018 | 0.8415 |
| Test Image 19 | 798.37 | 14.59 | 0.2784 | 0.0006 | 8.86 | 107.74 | 0.0237 | 55.14 | 0.8687 | 0.0086 | 0.7942 |
| Test Image 18 | 900.42 | 21.84 | 0.0531 | 0.0001 | 9.16 | 153.32 | 0.0122 | 31.09 | 0.5343 | 0.0005 | 0.5865 |
| Test Image 21 | 1125.39 | 26.01 | 0.0400 | 0.0001 | 9.89 | 140.98 | 0.0080 | 48.25 | 0.7583 | 0.0002 | 0.4731 |
| Test Image 17 | 1189.08 | 21.08 | 0.0944 | 0.0002 | 9.35 | 139.22 | 0.0123 | 47.90 | 0.7409 | 0.0008 | 0.7036 |
| Test Image 20 | 1373.55 | 26.37 | 0.0519 | 0.0002 | 9.00 | 202.34 | 0.0143 | 27.19 | 0.0709 | 0.0007 | 0.5321 |
| Test Image 23 | 1501.96 | 31.48 | 0.0300 | 0.0001 | 9.64 | 148.38 | 0.0089 | 31.81 | 0.2580 | 0.0002 | 0.3676 |
| Test Image 22 | 1706.43 | 33.37 | 0.0290 | 0.0001 | 9.82 | 117.83 | 0.0081 | 39.15 | 0.4434 | 0.0003 | 0.3525 |
| Test Image 24 | 2276.93 | 35.06 | 0.0424 | 0.0002 | 9.88 | 174.66 | 0.0135 | 57.28 | 0.6530 | 0.0089 | 0.3984 |

TABLE V.  TWENTY FOUR TEST IMAGES AND THEIR SECOND STATISTICS BASED ON DIAGONAL GLCM

| Image | Contrast | Dissimilarity | Homogeneity (Inverse Difference Moment) | Angular Second Moment (ASM) | Entropy | Mean (μ) | Energy | Standard Deviation (σ) | Correlation | Maximum Probability P(i,j) | Probability of Occurrence of Difference of 0 - 19 |
|---|---|---|---|---|---|---|---|---|---|---|---|
| Test Image 1 | 1.74 | 0.81 | 0.6715 | 0.0107 | 5.01 | 111.88 | 0.1034 | 10.47 | 0.9921 | 0.0307 | 1.0000 |
| Test Image 2 | 37.56 | 3.22 | 0.4356 | 0.0013 | 7.36 | 116.10 | 0.0366 | 30.63 | 0.9800 | 0.0068 | 0.9780 |
| Test Image 3 | 39.36 | 4.71 | 0.2064 | 0.0014 | 6.92 | 100.39 | 0.0373 | 10.71 | 0.8286 | 0.0035 | 0.9921 |
| Test Image 4 | 106.54 | 7.52 | 0.1434 | 0.0007 | 7.73 | 58.07 | 0.0261 | 18.08 | 0.8371 | 0.0018 | 0.9343 |
| Test Image 5 | 108.74 | 7.00 | 0.1953 | 0.0004 | 8.23 | 119.60 | 0.0209 | 28.71 | 0.9340 | 0.0018 | 0.9241 |
| Test Image 6 | 133.30 | 6.88 | 0.2346 | 0.0006 | 8.03 | 76.91 | 0.0239 | 25.52 | 0.8976 | 0.0029 | 0.9149 |
| Test Image 7 | 162.23 | 9.79 | 0.1068 | 0.0005 | 7.97 | 222.51 | 0.0218 | 16.82 | 0.7133 | 0.0023 | 0.8793 |
| Test Image 8 | 247.41 | 6.96 | 0.2904 | 0.0011 | 8.38 | 166.24 | 0.0336 | 60.46 | 0.9662 | 0.0270 | 0.9316 |
| Test Image 10 | 384.33 | 13.69 | 0.0962 | 0.0001 | 9.38 | 114.67 | 0.0108 | 49.83 | 0.9226 | 0.0011 | 0.7646 |
| Test Image 9 | 396.51 | 14.30 | 0.0802 | 0.0003 | 8.44 | 138.28 | 0.0181 | 20.06 | 0.5072 | 0.0009 | 0.7556 |
| Test Image 11 | 458.82 | 16.33 | 0.0654 | 0.0003 | 8.56 | 137.36 | 0.0167 | 19.48 | 0.3955 | 0.0008 | 0.6802 |
| Test Image 12 | 538.48 | 18.01 | 0.0570 | 0.0003 | 8.39 | 160.75 | 0.0181 | 17.31 | 0.1016 | 0.0009 | 0.6273 |
| Test Image 13 | 734.85 | 20.14 | 0.0665 | 0.0003 | 8.48 | 187.14 | 0.0186 | 29.84 | 0.5875 | 0.0014 | 0.6038 |
| Test Image 14 | 778.82 | 15.81 | 0.1052 | 0.0004 | 8.36 | 172.10 | 0.0206 | 26.49 | 0.4451 | 0.0014 | 0.7951 |
| Test Image 15 | 902.35 | 20.80 | 0.0641 | 0.0001 | 9.44 | 150.27 | 0.0114 | 52.41 | 0.8357 | 0.0031 | 0.6218 |
| Test Image 16 | 938.40 | 25.70 | 0.0346 | 0.0001 | 9.82 | 120.20 | 0.0080 | 50.91 | 0.8190 | 0.0002 | 0.4096 |
| Test Image 22 | 1184.56 | 27.47 | 0.0358 | 0.0001 | 9.72 | 117.84 | 0.0086 | 39.13 | 0.6132 | 0.0002 | 0.4313 |
| Test Image 18 | 1324.11 | 27.85 | 0.0397 | 0.0001 | 9.29 | 153.33 | 0.0112 | 31.09 | 0.3150 | 0.0006 | 0.4642 |
| Test Image 17 | 1344.91 | 23.37 | 0.0815 | 0.0001 | 9.44 | 139.19 | 0.0117 | 47.90 | 0.7069 | 0.0007 | 0.6572 |
| Test Image 20 | 1354.56 | 25.99 | 0.0510 | 0.0002 | 8.99 | 202.36 | 0.0143 | 27.17 | 0.0826 | 0.0008 | 0.5416 |
| Test Image 19 | 1533.77 | 23.12 | 0.1791 | 0.0003 | 9.30 | 107.98 | 0.0173 | 55.00 | 0.7465 | 0.0048 | 0.6762 |
| Test Image 21 | 1542.63 | 30.73 | 0.0340 | 0.0001 | 9.99 | 140.97 | 0.0076 | 48.24 | 0.6686 | 0.0002 | 0.4055 |
| Test Image 23 | 1724.51 | 33.88 | 0.0274 | 0.0001 | 9.66 | 148.40 | 0.0088 | 31.82 | 0.1484 | 0.0002 | 0.3375 |
| Test Image 24 | 3780.63 | 44.57 | 0.0392 | 0.0002 | 10.01 | 174.94 | 0.0135 | 57.23 | 0.4228 | 0.0102 | 0.3449 |